\documentclass[amsmath,amssymb,floatfix,preprint,showpacs,superscriptaddress,aps]{revtex4}

\usepackage{graphicx}
\usepackage{dcolumn}
\usepackage{bm}
\usepackage{amsmath}
\usepackage{amsthm}
\usepackage{xcolor}
\usepackage{multirow}

\usepackage[super]{nth} 

\begin{document}


\title{How many outbreaks before an epidemic?}

\author{Fabio Rapallo}
\email{fabio.rapallo@unige.it}
\affiliation{Department of Economics, Universit\`a di Genova, Italy}

\author{Enrico Scalas}
\email{enrico.scalas@uniroma1.it}
\homepage{https://corsidilaurea.uniroma1.it/it/users/enricoscalasuniroma1it}
\affiliation{Department of Statistical Sciences, Sapienza University of Rome, Italy}

\author{Pietro Terna}
\email{pietro.terna@unito.it}
\homepage{https://terna.to.it/}
\affiliation{Universit\`a di Torino, Italy, retired; Collegio Carlo Alberto, Torino, Italy}

\date{\today}



\begin{abstract}

In this work, we study the finite-population behaviour of the Reed–Frost epidemic model. Our analysis relies on the exact expression for the final epidemic size, replaced by Monte Carlo simulations in cases where the exact formula becomes numerically unstable. When the initial reproduction number is greater than a critical threshold, the distribution of the final size becomes bimodal. We therefore define the probabilities of small and large outbreaks, providing an intuitive answer to the question posed in the title through simple arguments based on the geometric distribution. Finally, an agent-based simulation confirms that the Reed-Frost model offers a good approximation in the case of the COVID-19 outbreak.

\end{abstract}

\maketitle

\section{Motivation} 
\label{motivation}

Stochastic models of epidemics show that in a fraction of cases an outbreak of an infectious disease dies out without leading to (exponential) growth of cases. An example of this behaviour is in reference \cite{ball19}, see in particular their figure 7. Irrespective of the detailed mechanism of the infection, let us assume than only two cases are possible (small outbreak or exponential growth as in \cite{ball19} or \cite{dilauro20}) and let us denote by $p$ the probability of developing an epidemic with (exponential) growth of infections. In general, this parameter will depend on the {\em initial conditions}, namely on the conditions under which the disease is introduced. An interesting question is: How many {\em small} independent outbreaks are needed before a full-fledged epidemic develops?

Assuming that no non-pharmaceutical measures are in place (if they are in place they may, and should, reduce $p$), and further assuming that introductions of the infectious disease are independent, we are left with a classical problem in probability theory: In a sequence of independent trials with two possible outcomes, what is the distribution of the time before the first ``success''? Here the success is the development of an exponentially growing epidemic. Let us call $T$ this time (which coincides with the number of trials), then the sought distribution is geometric
\begin{equation}
\label{geometric}
\mathbb{P}(T=k) = (1-p)^{k-1} p,
\end{equation}
for $k = 1, 2, \ldots$.
This is indeed the probability of the sequence $F, F, \ldots, F, S$, with $k-1$ $F$s and a final $S$, where $F$ stands for ``failure'' and $S$ stands for ``success''.
One has that the expected value of $T$ is given by:
\begin{equation}
\label{mean}
\mathbb{E} (T) = \frac{1}{p}
\end{equation}
and the variance of $T$ is
\begin{equation}
\label{variance}
\mathrm{Var}(T) = \frac{1-p}{p^2}.
\end{equation}
Therefore if, for example, $p=1/2$, then, in average, an epidemic will start after 2 independent introductions of the infectious disease, with $p=1/10$, an average of 10 independent introductions are needed and so on. 

These simple considerations can be helpful in the discussion around the origin of an infectious disease. However, what is missing from the above considerations is information on the physical-time structure of how the infectious disease is introduced. If $p=1/10$, for instance, and there are $500$ independent introductions of the disease in the very same day, a full-fledged epidemic will very likely develop on that day, perhaps with more than a single ``patient 1''. On the contrary, if the disease happens to be introduced at a slower pace with independent introductions of single cases separated by weeks, months or even years, it may take a long time for a full-fledged epidemic to develop even if $p$ is not very small.

The real world is however noisier and more complex than epidemic models, even if in their stochastic version. In order to deal with this complexity, we can study the outcome of agent-based simulations. In particular, to illustrate the above considerations, we use the program S.I.s.a.R., an agent-based model of the diffusion of COVID-19 using NetLogo, with Susceptible, Infected, symptomatic, asymptomatic, and Recovered people \cite{terna20} with code and further information publicly available at {\tt https://terna.to.it/simul/SIsaR.html}.
The model incorporates the medical perspective of one of its coauthors \cite{pescarmona20}.

However, to get a better insight into the origin of a full-fledged epidemics, we revert to simple mathematical models and make use of the classical Reed-Frost model for which several analytical results are available.

More details on S.I.s.a.R. and the Reed-Frost model are presented in the next section. The results of our study can be found in Section III. Section IV is then devoted to our final remark and to the applications of our considerations outside the field of mathematical epidemiology. As far as the latter point is concerned, we particularly focus on models for the introduction of {\em innovations}. Two of us already wrote a short paper in this topic (see \cite{rapallo2025}).

\section{The Models}
\label{model}

Here, we discuss results based on a realistic agent-based model developed for a specific geographic region and a specific disease (S.I.s.a.R.) and on a traditional mathematical epidemiological model: the Reed and Frost model. In this section, the reader finds a description of the two models.

\subsection{Agent based model}

S.I.s.a.R. is an agent-based model specifically designed to reproduce the diffusion of the COVID-19 using NetLogo, with Susceptible, Infected, symptomatic, asymptomatic, and Recovered people (hence the name S.I.s.a.R., it is indeed a modified agent-based SIR model where infected agents are categorized as symptomatic and asymptomatic). The model includes the structural data of Piedmont, an Italian region, but it can be calibrated for other areas following the Info sheet available online. It can reproduce the events of a realistic calendar (e.g., national or regional government decisions), via its script interpreter. The program is taking into account what was known on mechanisms of transmission of SARS-Cov-2 and on the novel coronavirus disease at the time of writing it during the acute phase of the pandemic. This can be modified as new information is emerging. For instance, a further modification of the algorithm could easily include a limited immunity period for recovered agents. However, {\em these details} (further specified below) {\em are largely irrelevant} for the purpose of the current paper, where we are just interested to see whether and when a full-fledged epidemic develops starting from a small number of infected agents introduced in the population.

We place two initial infected individuals in a population of 4350 individuals, in scale 1:1000 with Piedmont. The size of the initial infected group is out of scale: it is the smallest number ensuring the activation of the epidemic in a substantial number of cases. Initially infected people bypass the incubation period. For implausibility reasons, we never choose initial infected people among persons in nursing homes or hospitals.

We can set:

\begin{itemize}

\item minimum and maximum duration of the individual infection; 

\item the length of the incubation interval; 

\item the critical distance, as the radius of a circle affecting people which are in it, with a given probability; 

\item the correction of that probability due to the personal characteristics both of the active and the  passive agents; passive agents, as receivers, can be robust, regular, fragile, and extra fragile.

\end{itemize}

We have two main types of contagion: (a) within a radius, for people moving around, also if only temporary present in a house/factory/nursing home/hospital (in schools we just have students and teachers); (b) in a given space (room or apartment) for people resident in their home or in a hospital or in a nursing home or being in school or in a working environment.

People in hospitals and nursing homes can be infected in two ways: (a) and (b). While people are at school, they can only pass the disease to people in the same classroom, where only teachers and students are present, so this is a third infection mechanism (c). One should remark that workplaces are open to agents, as clients, vendors, suppliers, external workers can go there whereas schools are mainly reserved to students and school operators and are less affected by contacts with agents of other type.

All agents have their home, inside a city or a town. The agents also have a regular place (RP) where they act and interact, moving around. These positions can be interpreted as free time elective places. Students and teachers, when we activate the school, have both RPs and the schools; healthcare operators have both RPs and hospital or nursing homes; finally, workers have both RPs and working places. In each day (or tick of the model) we simulated realistic sequences of actions.

\subsection{The model by Reed and Frost}

Agent-based models can be a useful tool for planning and scenario building, but they might be quite difficult to analyse using mathematical tools in a rigorous way. For this reason, we resort to a classical model, introduced by Reed and Frost in the 1920s, in order to clarify the onset of full-fledged epidemics.

The Reed-Frost model is a Markov chain in discrete time with discrete state space. The population at time step $t$ is divided into susceptibles $S_t$, infectives $I_t$ and recovered or removed $N-I_t-S_t$, where $N=S_0+I_0$ is the population size. At time $t=0$ we have no recovered. In all scenarios, we start with a population with $S_0=s_0$ susceptibles and we consider two possible initial $I_0$, i.e., $I_0=i_0=1$ infective and $I_0=i_0=2$ infectives. In the Reed-Frost model, all individuals in the population are connected and each edge has a common probability of contagion $\tilde{p} \in (0,1)$. Thus, the initial reproduction rate is ${R}_0=\tilde{p} s_0$. At each step, given $S_t=s_t$ and $I_t=i_t$, the evolution of the epidemic is ruled at time $t+1$ by the equations below
\[
S_{t+1} \sim \mathrm{Bin}\left( s_t , (1-\tilde{p})^{i_t} \right), \qquad \qquad I_{t+1}=s_t-S_{t+1}.
\]
The duration of the epidemics is
\begin{equation}\label{epduration}
\tau = \min \left\{ t \ | \ I_t=0 \right\}
\end{equation}
and the epidemic size (it could also be referred to as the {\em final} size of the epidemics) is therefore:
\begin{equation}\label{epsize}
W = \sum_{t=0}^\tau I_t = N-S_{\tau}.
\end{equation}
As we shall see below, an exact formula for the size of the epidemic $T$ is available for the Reed-Frost model on the fully connected graph. As a final remark, we notice that the Reed-Frost model is related to bond percolation and the size of the epidemic coincides with the size of the giant component of bond percolation on the fully connected graph. To the best of our knowledge, bond percolation was introduced in 1957 by Broadbent and Hammersley in Section 3 of \cite{broadbent1957}. In rather abstract terms, it can be described as follows: Let $G = (V,E)$ denote a graph with $V$ being the set of vertices and $E$ the set of edges (bonds). Let us associate to each edge $e \in E$ a probability $p(e) \in [0,1]$. The bond percolation graph is a subgraph $G_{p(e)} = (V, E_{p(e)})$ where $e \in E_{p(e)}$ with probability $p(e)$. If we set $p(e) = \tilde{p}$ for every edge $e$ in the graph, we can see that the probability distribution of the final size of the epidemics $T$ is the same as the probability distribution of the so-called giant component in bond percolation \cite{newman02}. The equivalence between bond percolation and the Reed-Frost model is an established fact in mathematics/probability theory (e.g. see section 5.3 in \cite{durrett07}), physics \cite{newman02,miller16}, as well as in computer science (in particular see appendix A.2 in \cite{becchetti22}). As written by Miller in \cite{miller16}: {\em  This percolation equivalence is based on the fact that an edge either exists or does not in percolation, while in disease spread if the edge transmits, the receiving node becomes infected.} Indeed, percolation models were originally motivated also by epidemiology (see \cite{broadbent1957,frisch1963}).

We decided to use the Reed-Frost model for two main reasons: first, to have a flexible tool for simulations and computations because of its simple structure; second, its use in similar applications in the recent literature, see for instance \cite{miller16, govindakutty2024, rapallo2025}.

\section{Results}

As mentioned in the previous section, in all the simulations of the agent-based model described in the next two subsections, we consider an initial state where two asymptomatic infected individuals are placed in the world randomly. Every simulation is independent from the other ones and represents a possible scenario of an experiment. We present the result of two experiments. In the first one, the epidemic develops without non-pharmaceutical containment measures. In the second experiment, we consider successive containment measures following the calendar mentioned in the previous section. The sequence of main events is reported in Table \ref{tab:events}. The results for the Reed-Frost models are presented in the third subsection.

\begin{table}[ht]
\centering
\begin{tabular}{|l|l|}
\hline
Day & Event \\ \hline 
Day~1 & Conventionally, in the model the epidemic starts on Feb \nth{3}, 2020 \\
Day~17 & Due to carnival holidays, schools close \\
Day~20 & Piedmont Region first warning, with the prohibition of crowd gatherings \\
Day~35 & Limitation of movement outside local areas \\
Day~38 & Full lockdown on March \nth{11} \\
Day~49 & Almost total blockage of non-essential production activities \\
Day~84 & Reduction of the limitations \\
Day~106 & Elimination of a large part of the restrictions; schools always inactive\\ 
\hline
\end{tabular}
\caption{Calendar of events.}
    \label{tab:events}
\end{table}

\subsection{Agent-based model: Unrestricted epidemic}

\begin{figure}
\includegraphics[width=0.95\textwidth]{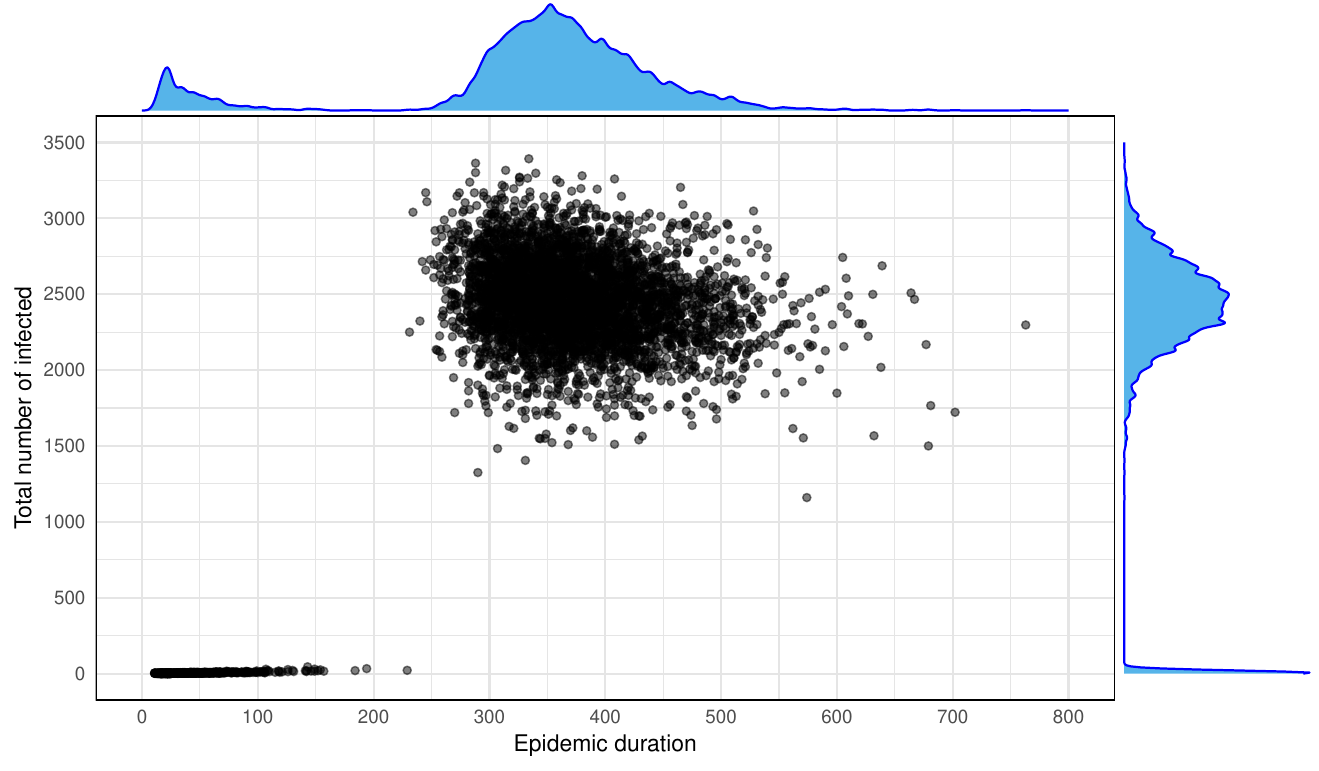}
\caption{Scatterplot of the final total number of infected individuals of all categories versus the duration of the epidemic without containment measures.} \label{fig:unrestr}
\end{figure}

We simulate 5000 independent scenarios of an unrestricted epidemic with S.I.s.a.R.. The simulation stops when the epidemic dies out and no symptomatic or asymptomatic infectious agents remain. In the case of a full-fledged epidemic, this is a situation in which so-called {\em herd immunity} is reached and then the epidemic dies out naturally. On the contrary, for small outbreaks chance leads to a stop of the epidemics for lack of contacts. As remarked in the previous section, a recent study seems to point to a short-term immunity to SARS-Cov-2 \cite{seow20} in agreement with what happens for other human coronaviruses and the model might need to be modified in order to take this phenomenon into account. In Figure \ref{fig:unrestr}, we present the scatterplot of the final total number of infected individuals of all categories versus the duration of the epidemic. In the same figure, also a kernel density estimation of the marginal distributions is shown.

The marginal distribution of the final total number of infected individuals can be compared with figure 7 in reference \cite{ball19} and it is compatible with the discussion in Section \ref{motivation} with $p=0.9$. Small and large outbreaks give rise to a clear bimodal scatterplot. One can also see that the largest epidemics ends with $3100$ infected agents ($71.3$\% of the total population), and the mode of the mass for full-fledged epidemics is $2450$ ($56$\% of the total population). These estimates do not seem in disagreement with what can be observed in epidemic models in the presence of population heterogeneity \cite{britton20,gomes20}, but they are not as optimistic as the results in those papers. Smaller outbreaks are also shorter, but the overall relationship between size and duration of the epidemic is not linear as shown in the scatter plot of Figure \ref{fig:unrestr}.

\subsection{Agent-based model: Epidemic with containment measures}

\begin{figure}
\includegraphics[width=0.95\textwidth]{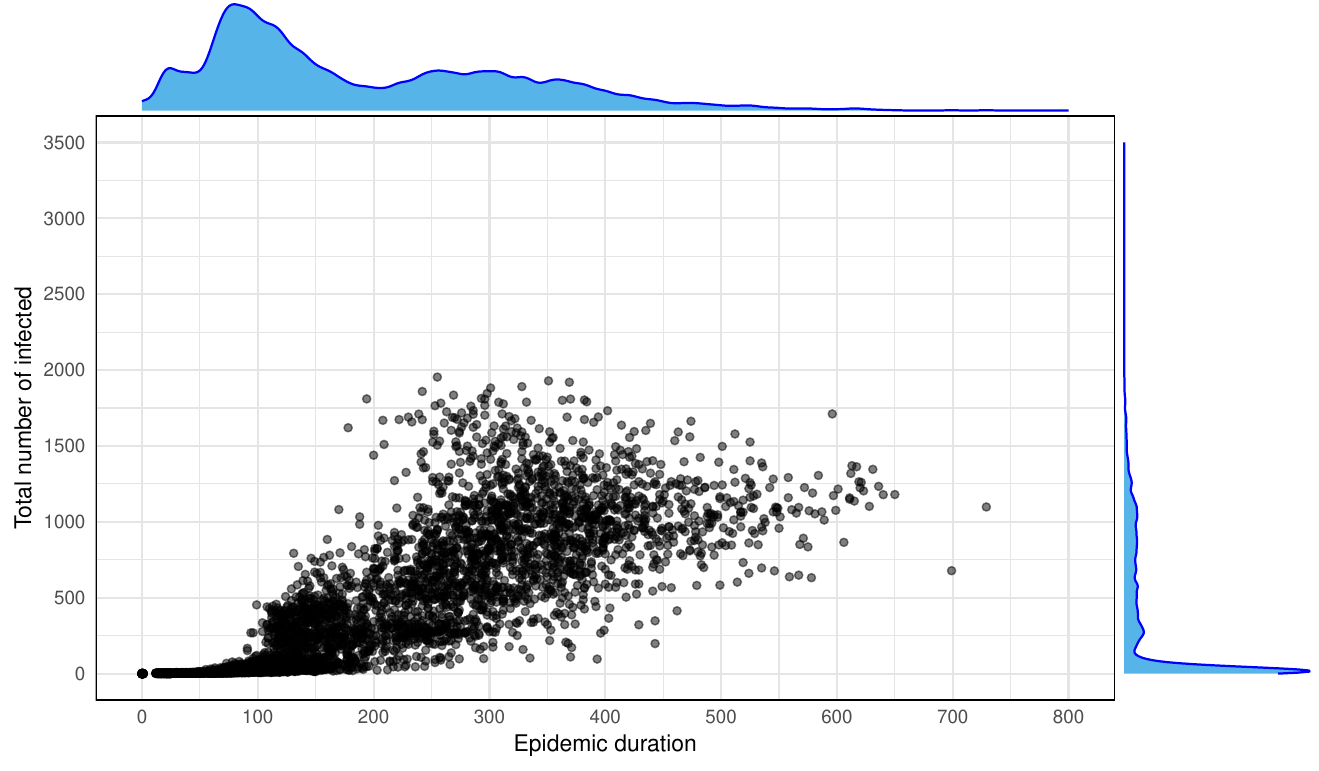}
\caption{Scatterplot of the final total number of infected individuals of all categories versus the duration of the epidemic with containment measures.} \label{fig:restr}
\end{figure}

In the presence of containment measures, the picture becomes quite interesting. We run 5000 simulations with the same initial conditions as before and, in Figure \ref{fig:restr}, we show the scatterplot of the final total number of infected individuals of all categories versus the duration of the epidemic, analogously to Figure \ref{fig:unrestr}.

In this case, there are epidemics of all sizes up to a maximum one which is smaller than the typical size of the unrestricted outbreaks, namely the maximum size in this case is around $1800$ infected agents ($41.4$\% of the total population). The probability of such an extreme event appears to be quite small and the majority of scenarios have a total number of infected agents below $28.7$\% of the total population. Also the probability of small outbreaks is higher than in the unrestricted case.

Even if also in this case, the distribution of durations is bimodal, its expected value appears to be remarkably smaller than in the unrestricted case. The bivariate scatterplot shows again two clusters (small and large outbreaks) as in the unrestricted case in Figure \ref{fig:restr}, but the separation between the two clusters is now less perceptible. Finally, as in the previous case, the largest epidemic is not the longest even if there is a positive correlation between size and duration.

\subsection{Simulations of the Reed-Frost model}

In this section we present simulations on the simple Reed-Frost model in order to evaluate the probabilities of small and large outbreaks. We consider several scenarios characterised by different population sizes and different probabilities of transmission and we focus on the final size of the epidemics $W$. 
$W$ has been studied asymptotically in, e.g., \cite{martinlof1} and \cite{martinlof2}. In our finite-population approach, the final size, defined in Eq.~\eqref{epsize}, is a random variable and its distribution is known exactly, see \cite{lefevre90,picard90,ball99}: 
\begin{equation}
\label{finsize}
{\mathbb P}(W=k+1) = \frac {s_0!} {(s_0-k)!} (1-\tilde{p})^{(1+k)(s_0-k)} G_k(1 \ | \ U) \qquad (k \ge 0)
\end{equation}
where $U$ is the sequence $((1-\tilde{p}), (1-\tilde{p})^2, \ldots, (1-\tilde{p})^{(1+s_0)}, 0 ,0 , \ldots)$ and $G_k(1 \ | \ U)$ is defined recursively in terms of Goncharov polynomials by:
\[
G_0(1 \ | \ U)=1 \qquad \qquad G_k(1 \ | \ U)=\frac 1 {k!} - \sum_{j=0}^{k-1} \frac {u_j^{k-j}} {(k-j)!} G_j(1 \ | \ U) \qquad (k\ge 1).
\]
The formula is not numerically stable when the population grows (see section 3.1 in \cite{britton10}, for example), even if one uses high precision arithmetic through the \texttt{R} package \texttt{Rmpfr} \cite{Rmpfrmanual}. As a consequence, we use both the exact formula and the Monte Carlo estimate for small values of $s_0$ and only Monte Carlo estimation for larger values of $s_0$. Monte Carlo estimates of the probabilities are based on $5,000$ trajectories, and $2,000$ replicates are used to estimate the standard deviations. All the simulations are carried out using \texttt{R} version 4.5.1 \cite{Rproject}. Although approximate results exist in the case of large populations \cite{martinlof1,martinlof2,britton10}, here we prefer to use Monte Carlo simulations as we explore values of $s_0 \leq 512$ still far from the asymptotic regime.

\begin{figure}
\includegraphics[width=\textwidth]{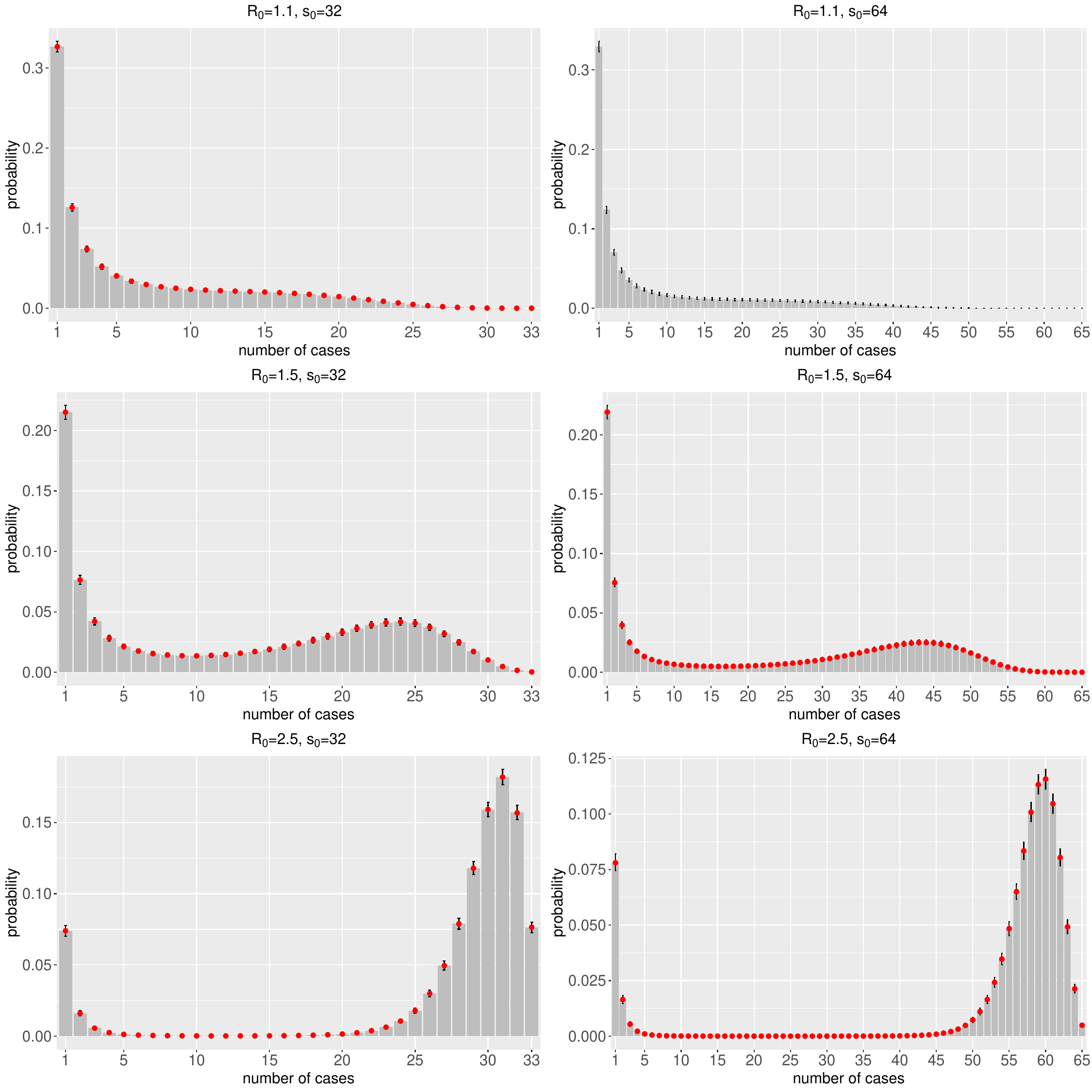}
\caption{Six distributions of the final epidemic size. The grey rectangles are the Monte Carlo estimates, the segments are the sd, while red dots represent the exact probabilities. When the exact formula is not stable, only the Monte Carlo results are shown.}
\label{fig_6plots}
\end{figure}

For the considered values of $s_0$ and $R_0$, the final epidemic size distributions are shown in Fig. \ref{fig_6plots}.

Looking at the distributions in Fig.~\ref{fig_6plots}, we observe a common pattern for $R_0=1.5$ and $R_0=2.5$: There are two peaks, one peak of small outbreaks where few individuals become infective and another peak of large outbreaks where most of the individuals in the population become infective. On the other hand, the top panels in Fig.~\ref{fig_6plots}, for $R_0=1.1$, do not exhibit a bimodal behaviour; we shall come back on this issue at the end of this section. Now, we concentrate on scenarios with a bimodal density, where it is easy to separate a probability of small outbreak and a probability of large outbreak. We then define the probability of small outbreak by means of the following rule: (i) first, we determine the relative minimum $t_m$ of the distribution between the two peaks, as the separator between small and large outbreaks; (ii) then, the probability of small outbreak is defined as the sum
\begin{equation}\label{probsmall}
q_s = \sum_{k=0}^{t_m-1} \mathbb{P}(W=k) \, .
\end{equation}
The quantity $p_s = 1-q_s$ is a proxy of the probability $p$ discussed in the introduction.

In the Monte Carlo approach, the minimum is determined after a standard smoothing through a rolling mean (for population size smaller than 100 people) or a kernel smoothing (for population size larger than 100 people), and looking for sign changes in the differences. If more than one local minimum is found due to statistical variability in the density estimation, we take the (rounded) median of the points of local minimum. Then, the estimate $\hat{q}_s$ of $q_s$ is obtained.

The results for different scenarios are reported in Table \ref{tab:1} (for scenarios with small populations where exact probabilities and Monte Carlo estimates can be used) and in Table \ref{tab:2} (for scenarios where only Monte Carlo estimates are reliable). From these results, we can see that the probability of small outbreak is greater than $0.20$ when $R_0\le 2$ in the case $i_0=1$ (except for the case $s_0=16$), and when $R_0 \le 1.4$ in the case $i_0=2$. In the other direction, the probability of small outbreak is less than $0.01$ when $R_0\ge 2.6$ in the case of two initial infectives. In particular, the results with $i_0=1$  show that, even in cases of epidemics with a moderate initial reproduction number, still there are non-negligible probabilities of small outbreak and such values tend to stabilise as the population size grows. 

\begin{table}[ht]
\centering
\begin{footnotesize}
\renewcommand\arraystretch{0.9}
\begin{tabular}{c|c|ccc|ccc}
\hline
& & \multicolumn{3}{c|}{$i_0=1$} & \multicolumn{3}{|c}{$i_0=2$} \\ \hline
 ${R}_0$  & $s_0$ & exact $q_s$ & mean($\hat{q}_s$) & sd($\hat{q}_s)$ & exact $q_s$ & mean($\hat{q}_s)$ & sd($\hat{q}_s)$ \\ \hline
1.2 & 16 & 0.6255 & 0.6547 & 0.0454 & 0.2675 & 0.2765 & 0.0507 \\
1.2 & 32 & 0.7030 & 0.7321 & 0.0331 & 0.3918 & 0.4682 & 0.0640 \\
1.2 & 64 & NS & 0.8237 & 0.0581 & NS & 0.6255 & 0.0925 \\
1.4 & 16 & 0.4565 & 0.4635 & 0.0263 & 0.1341 & 0.1521 & 0.0232 \\
1.4 & 32 & 0.5112 & 0.5289 & 0.0390 & 0.2233 & 0.2406 & 0.0360 \\
1.4 & 64 & NS & 0.5549 & 0.0427 & NS & 0.2962 & 0.0492 \\
1.6 & 16 & 0.3501 & 0.3463 & 0.0157 & 0.0829 & 0.0931 & 0.0155 \\
1.6 & 32 & 0.3840 & 0.3847 & 0.0155 & 0.1323 & 0.1404 & 0.0170 \\
1.6 & 64 & 0.3886 & 0.3932 & 0.0166 & 0.1542 & 0.1567 & 0.0115 \\
1.8 & 16 & 0.2639 & 0.2619 & 0.0117 & 0.0612 & 0.0580 & 0.0082 \\
1.8 & 32 & 0.2841 & 0.2855 & 0.0087 & 0.0807 & 0.0822 & 0.0079 \\
1.8 & 64 & 0.2835 & 0.2848 & 0.0070 & 0.0837 & 0.0848 & 0.0045 \\
2 & 16 & 0.1970 & 0.1967 & 0.0084 & 0.0357 & 0.0352 & 0.0051 \\
2 & 32 & 0.2098 & 0.2121 & 0.0074 & 0.0470 & 0.0469 & 0.0040 \\
2 & 64 & 0.2091 & 0.2091 & 0.0056 & 0.0457 & 0.0460 & 0.0030 \\
2.2 & 16 & 0.1463 & 0.1469 & 0.0064 & 0.0205 & 0.0211 & 0.0031 \\
2.2 & 32 & 0.1571 & 0.1574 & 0.0056 & 0.0264 & 0.0264 & 0.0025 \\
2.2 & 64 & 0.1565 & 0.1568 & 0.0050 & 0.0254 & 0.0255 & 0.0022 \\
2.4 & 16 & 0.1086 & 0.1097 & 0.0051 & 0.0117 & 0.0123 & 0.0020 \\
2.4 & 32 & 0.1175 & 0.1178 & 0.0046 & 0.0148 & 0.0149 & 0.0018 \\
2.4 & 64 & 0.1191 & 0.1193 & 0.0045 & 0.0146 & 0.0146 & 0.0018 \\
2.6 & 16 & 0.0808 & 0.0818 & 0.0041 & 0.0071 & 0.0071 & 0.0014 \\
2.6 & 32 & 0.0891 & 0.0893 & 0.0041 & 0.0084 & 0.0084 & 0.0013 \\
2.6 & 64 & 0.0918 & 0.0919 & 0.0041 & 0.0086 & 0.0086 & 0.0013 \\
2.8 & 16 & 0.0610 & 0.0608 & 0.0035 & 0.0040 & 0.0049 & 0.0010 \\
2.8 & 32 & 0.0681 & 0.0682 & 0.0035 & 0.0049 & 0.0049 & 0.0010 \\
2.8 & 64 & 0.0714 & 0.0715 & 0.0036 & 0.0052 & 0.0052 & 0.0010 \\
3 & 16 & 0.0454 & 0.0454 & 0.0030 & 0.0022 & 0.0022 & 0.0007 \\
3 & 32 & 0.0524 & 0.0524 & 0.0031 & 0.0029 & 0.0029 & 0.0008 \\
3 & 64 & 0.0559 & 0.0558 & 0.0032 & 0.0032 & 0.0032 & 0.0008 \\
\hline
\end{tabular}
\end{footnotesize}
    \caption{Probability of small outbreak $q_s$ for scenarios with both exact and Monte Carlo computations. (NS is displayed when the exact formula is numerically not stable.)}
    \label{tab:1}
\end{table}

\begin{table}[ht]
\centering
\begin{footnotesize}
\renewcommand\arraystretch{0.9}
\begin{tabular}{c|c|cc|cc}
\hline
& & \multicolumn{2}{c|}{$i_0=1$} & \multicolumn{2}{|c}{$i_0=2$} \\ \hline
 ${R}_0$  & $s_0$ & mean($\hat{q}_s$) & sd($\hat{q}_s)$ & mean($\hat{q}_s)$ & sd($\hat{q}_s)$ \\ \hline
1.2 & 128 & 0.7578 & 0.0319 & 0.5438 & 0.0401 \\
1.2 & 256 & 0.7459 & 0.0305 & 0.5407 & 0.0222 \\
1.2 & 512 & 0.7293 & 0.0208 & 0.5324 & 0.0182 \\
1.4 & 128 & 0.5330 & 0.0090 & 0.2874 & 0.0091 \\
1.4 & 256 & 0.5192 & 0.0075 & 0.2748 & 0.0071 \\
1.4 & 512 & 0.5046 & 0.0072 & 0.2578 & 0.0061 \\
1.6 & 128 & 0.3805 & 0.0071 & 0.1497 & 0.0067 \\
1.6 & 256 & 0.3684 & 0.0068 & 0.1384 & 0.0049 \\
1.6 & 512 & 0.3624 & 0.0067 & 0.1320 & 0.0048 \\
1.8 & 128 & 0.2757 & 0.0062 & 0.0784 & 0.0038 \\
1.8 & 256 & 0.2707 & 0.0063 & 0.0740 & 0.0037 \\
1.8 & 512 & 0.2689 & 0.0063 & 0.0726 & 0.0037 \\
2 & 128 & 0.2051 & 0.0058 & 0.0431 & 0.0029 \\
2 & 256 & 0.2041 & 0.0057 & 0.0420 & 0.0029 \\
2 & 512 & 0.2035 & 0.0056 & 0.0416 & 0.0028 \\
2.2 & 128 & 0.1561 & 0.0052 & 0.0246 & 0.0022 \\
2.2 & 256 & 0.1561 & 0.0052 & 0.0245 & 0.0022 \\
2.2 & 512 & 0.1562 & 0.0051 & 0.0245 & 0.0022 \\
2.4 & 128 & 0.1201 & 0.0046 & 0.0145 & 0.0017 \\
2.4 & 256 & 0.1207 & 0.0047 & 0.0146 & 0.0017 \\
2.4 & 512 & 0.1210 & 0.0048 & 0.0147 & 0.0017 \\
2.6 & 128 & 0.0934 & 0.0042 & 0.0089 & 0.0013 \\
2.6 & 256 & 0.0943 & 0.0042 & 0.0090 & 0.0013 \\
2.6 & 512 & 0.0947 & 0.0042 & 0.0089 & 0.0014 \\
2.8 & 128 & 0.0730 & 0.0037 & 0.0054 & 0.0011 \\
2.8 & 256 & 0.0742 & 0.0038 & 0.0055 & 0.0010 \\
2.8 & 512 & 0.0746 & 0.0037 & 0.0056 & 0.0011 \\
3 & 128 & 0.0578 & 0.0033 & 0.0033 & 0.0008 \\
3 & 256 & 0.0585 & 0.0034 & 0.0035 & 0.0008 \\
3 & 512 & 0.0590 & 0.0034 & 0.0035 & 0.0008 \\
\hline
\end{tabular}
\end{footnotesize}
\caption{Probability of small outbreak $q_s$ for scenarios with Monte Carlo computations only.}
    \label{tab:2}
\end{table}

\begin{figure}
\begin{center}
\includegraphics[width=\textwidth]{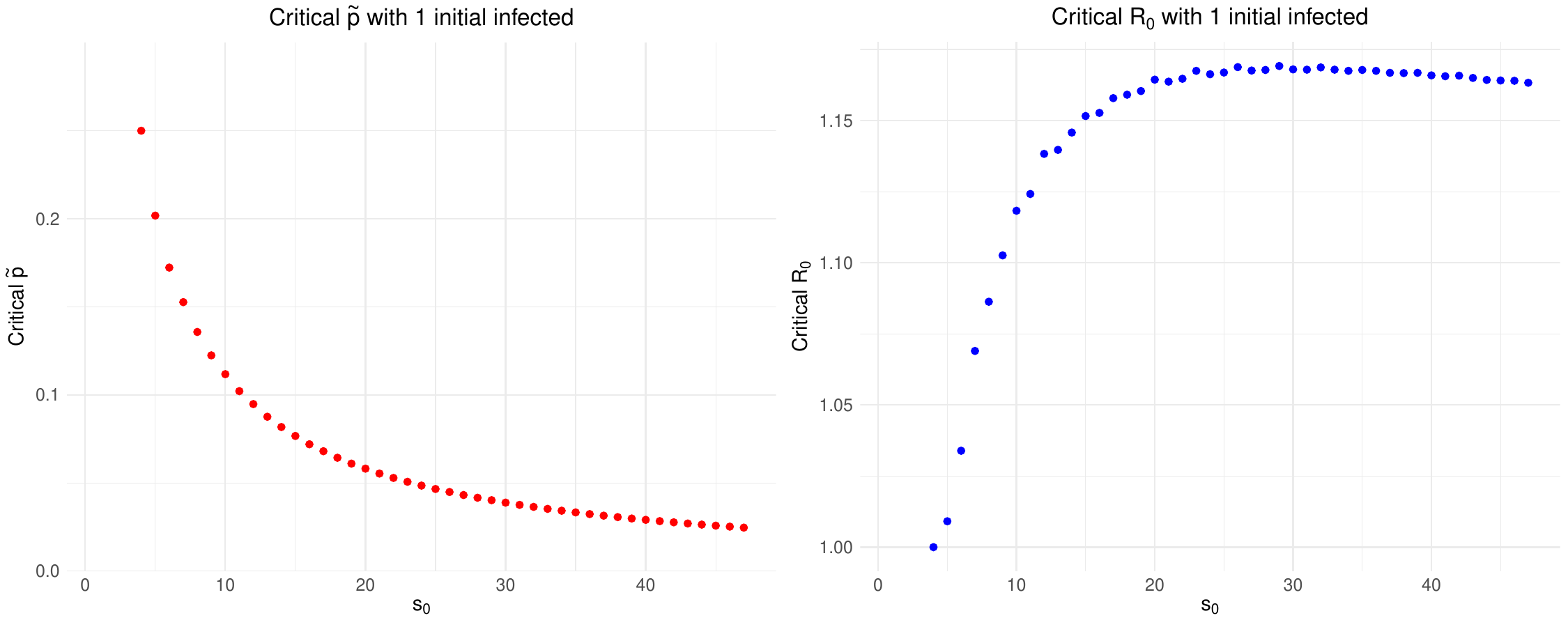}
\end{center}
\caption{Critical values of $\tilde{p}$ (left panel) and $R_0$ (right panel) with $i_0=1$.} \label{fig:crittildepR0}
\end{figure}

When the probability of infection is too small, the second peak does not appear in the density, leading to a monotone decreasing density function, as shown in the top panels of Fig.~\ref{fig_6plots}, where the initial reproduction number is $R_0=1.1$. In this case, the definition of probability of small and large outbreak in Eq.~\eqref{probsmall} cannot be applied. Thus, to further investigate this issue, using the exact formula, we have considered the {\em critical value} of the infection probability $\tilde{p}$ (or the critical value of the initial reproduction number ${R}_0$). In this finite-size analysis, the critical value is defined as the minimum value of the parameter for which there are two modes in the density of the final epidemic size when $i_0=1$. For this study, we limit the population size to those cases where the exact formula yields stable results. In Figure \ref{fig:crittildepR0}, all the critical values are presented, after computing them using the exact formula for the final size Eq. \eqref{finsize}. Heuristically, one expects that the critical value for this transition to occur can be found by setting $\tilde{p}_c s_0 \approx 1$ coinciding with a critical value of $R_0$ close to 1 and leading to $\tilde{p}_c \approx 1/s_0$, which is confirmed by the plot. However, the reader should keep in mind that, due to the finite size and to the definition of $R_0$, the critical value of $R_0$ is close but not equal to 1. In fact, while the critical value of $\tilde{p}$ decreases with the population size, the critical value of $R_0$ has a different behaviour. After an initial region affected by small-sample effects, it exhibits a near-constant behaviour. For instance, with $s_0=16$ we have a critical value $R_0=1.1528$, with $s_0=32$ we have a critical $R_0=1.1688$, and with $s_0=45$ we have a critical $R_0=1.1641$. We expect that the critical $R_0$ converges to 1 for large system sizes. This is illustrated in Table \ref{around-one}, where as a function of $R_0$, we report the median final size, the 90th percentile for the final size, and the probability that the epidemics hits less than $5\%$ of susceptible individuals for $s_0+1 = N = 1,000$ and $10,000$ individuals and one initial infective.

\begin{table}[ht]
\centering
\begin{footnotesize}
\renewcommand\arraystretch{0.9}
\begin{tabular}{cccc|cccc}
\hline
\multicolumn{4}{c|}{$N = 1{,}000$} & \multicolumn{4}{c}{$N = 10{,}000$} \\
\hline
$R_0$ & Median & Q90 & Prob(W $<$ $5\%$ pop.) & 
$R_0$ & Median & Q90 & Prob(W $<$ $5\%$ pop.) \\
\hline
1.10 & 2 & 188 & 0.8017 & 1.10 & 2 & 1701 & 0.8349 \\
1.08 & 2 & 149 & 0.8236 & 1.08 & 2 & 1179 & 0.8687 \\
1.06 & 2 & 113 & 0.8444 & 1.06 & 2 & 456  & 0.9023 \\
1.04 & 2 &  84 & 0.8644 & 1.04 & 2 & 165  & 0.9328 \\
1.02 & 2 &  63 & 0.8831 & 1.02 & 2 &  91  & 0.9586 \\
1.00 & 1 &  49 & 0.9007 & 1.00 & 1 &  61  & 0.9773 \\
0.98 & 1 &  39 & 0.9169 & 0.98 & 1 &  45  & 0.9892 \\
0.96 & 1 &  31 & 0.9317 & 0.96 & 1 &  35  & 0.9957 \\
0.94 & 1 &  26 & 0.9443 & 0.94 & 1 &  28  & 0.9985 \\
0.92 & 1 &  22 & 0.9556 & 0.92 & 1 &  23  & 0.9996 \\
0.90 & 1 &  19 & 0.9653 & 0.90 & 1 &  20  & 0.9999 \\
\hline
\end{tabular}
\end{footnotesize}
\caption{Behaviour of the final size $W$ around $R_0=1$ for two large values of $N$ by Monte Carlo simulation.} \label{around-one}
\end{table}

\section{Final remarks and discussion}

In this note, we have used a SIR-like agent-based model including asymptomatic and symptomatic infectious agents called S.I.s.a.R. and written in NetLogo to corroborate an elementary probabilistic model for the number of independent trials needed for the development of a full-fledged epidemic after the introduction of a new infectious disease in a fully susceptible population. The qualitative behaviour of the agent-based model is the same as the qualitative behaviour of probabilistic SIR models such as \cite{ball19}. In the absence of non-pharmaceutical prevention measures, there are either realisations of the process (runs of the simulation) in which the outbreak dies out before reaching the stage of initial exponential growth or realisations in which exponential growth is reached and a full-fledged epidemic develops. On the contrary, when specific non-pharmaceutical measures are put in place according to a given calendar, one can observe finite probabilistic mass for all epidemic sizes up to the maximum. We are also able to study the duration of the epidemic and how it correlates with the size.

To better understand the emergence of the behaviour described above, we have studied the finite and small-size behaviour of the model introduced by Reed and Frost. Thanks to the existence of an exact formula for the final size of the epidemic (Eq. \eqref{finsize}) we have been able to validate Monte Carlo simulations and we have performed a detailed Monte Carlo study of a transition of the final size distribution from unimodal to bimodal. In the bimodal case, we have been able to estimate the probability of developing a full-fledged epidemic in a small population of susceptible individuals when 1 and 2 infective individuals are introduced. Incidentally, the presence of this transition is not restricted to the fully-connected graph case we consider here \cite{dilauro20}.

From the point of view of applications, SIR-like models have been used to describe the diffusion of rumours in human populations (see for example \cite{zapperi25}) and they can be adapted to describe the diffusion of ideas (see for example \cite{bettencourt06} on the diffusion of the use of Feynman diagrams in theoretical physics).

This is based on an analogy between the spread of ideas and the spread of infectious diseases that was quite popular in the second half of the 20th Century. In a recent paper of ours \cite{rapallo2025}, we mentioned a 1964 paper by Goffman and Newill as one of the first sources we could find \cite{goffmann1964}. It is tempting to exploit this analogy and consider the simple model outlined in the introduction as a description of the number of introductions of a new idea or concept needed before it becomes successful. For example, the history science is full of instances in which a very successful model, theory, idea or method had already been introduced in the past at a time in which nobody really cared. In \cite{rapallo2025}, we mention two cases: the discovery of penicillin by Vincenzo Tiberio well before Fleming \cite{percacciante} and the theory of option pricing developed by Vinzenz Bronzin well before Black and Scholes \cite{zimmermann}. At this stage, this is just a nice conjecture, but we hope to either corroborate or falsify it soon.











\end{document}